\documentclass[aip,jcp,preprint]{revtex4-1}
\usepackage{graphicx}
\usepackage{pstricks}
\usepackage{amsmath}
\usepackage{amssymb}
\usepackage{bm}
\newcommand{\ii}{\ensuremath{\textrm{i}}}
\newcommand{\ud}{\ensuremath{\textrm{d}}}
\newcommand{\tr}{\ensuremath{\textrm{Tr}}}
\newcommand{\cm}{\ensuremath{\textrm{cm}^{-1}}}

\newcommand{\ket}[1]{\ensuremath{\left| #1 \right\rangle}}
\DeclareMathOperator{\sgn}{sgn}

\begin{document}
\title{Linear and non-linear infrared response of one-dimensional vibrational Holstein polarons in the anti-adiabatic limit: optical and acoustical phonon models}
\author{Cyril Falvo}
\email{cyril.falvo@u-psud.fr}
\affiliation{Institut des Sciences Mol\'eculaires d'Orsay (ISMO), CNRS, Univ. Paris-Sud, Universit\'e Paris-Saclay, 91405 Orsay, France}
\affiliation{Univ. Grenoble Alpes, CNRS, LIPhy, 38000 Grenoble, France}

\begin{abstract}
The theory of linear and non-linear infrared response of vibrational Holstein polarons in one-dimensional lattices is presented in order to identify the spectral signatures of self-trapping phenomena. Using a canonical transformation the optical response is computed from the small polaron point of view which is valid in the anti-adiabatic limit. Two types of phonon baths are considered: optical phonons and acoustical phonons, and simple expressions are derived for the infrared response. It is shown that for the case of optical phonons, the linear response can directly probe the polaron density of states.  The model is used to interpret the experimental spectrum of crystaline actetanilide in the C$=$O range. For the case of acoustical phonons, it is shown that two bound states can be observed in the two-dimensional infrared spectrum at low temperature. At high temperature, analysis of the time-dependence of the two-dimensional infrared spectrum indicates that bath mediated correlations slow down spectral diffusion. The model is used to interpret the experimental linear-spectroscopy of model $\alpha$-helix and $\beta$-sheet polypeptides. This work shows that the Davydov Hamiltonian cannot explain the observations in the NH stretching range.
\end{abstract}

\maketitle

\section{Introduction}
The dynamics of electronic or vibrational excitons in quasi one-dimensional lattices has been an open topic of research for the past 60 years.\cite{Mahan:1981aa,Holstein:1959mz,Holstein:1959kx} The interplay between exciton delocalization and  coupling with the lattice vibrations results in the self-trapping phenomena, i.e. the formation of a polaron. A polaron usually refers to a quasi-particle that comprises the exciton and the lattice deformation created by the exciton which modifies its dynamics.\cite{Mahan:1981aa,Holstein:1959mz,Holstein:1959kx} The concept of self-trapping in one-dimensional lattices has a large number of applications for example in molecular aggregates,\cite{Spano:2010kq,Huynh:2013fk,Lu:1991fv,Sun:2015fk,Chorosajev:2014zj,Chorosajev:2016ap} conjugated polymers,\cite{Yamagata:2014rg,Barford:2014dq} halogen-bridged metal complexes,\cite{Okamoto:1992ye} molecular crystals\cite{Fillaux:1981uq,Barthes:1998fk,Herrebout:2001uq,Careri:1983vn,Careri:1984fd,Eilbeck:1984kx,Alexander:1986yu,Edler:2002uq,Edler:2002kc,Edler:2003vg,Hamm:2006uq} and biological macromolecules.\cite{Davydov:1973qr,Davydov:1985aa,Scott:1982gn,Scott:1992ty,Edler:2004qy,Edler:2005gf,Hamm:2009kx,Brown:1989ve,Ivic:1997lq,Pouthier:2003fj,Pouthier:2004fk,Falvo:2005ul,Falvo:2005qf,Falvo:2005ve,Tsivlin:2006qf,Tsivlin:2006ve,Tsivlin:2007qf,Bodis:2009uq,Cruzeiro:2009oz,Goj:2011vn} 
Vibrational excitons emerge in molecular crystal or within biological macromolecules by the delocalization of high-frequency vibrations through dipole-dipole interactions. This is the case for example, in $\alpha$-helix polypeptides where the amide-I band corresponds to the delocalization of the C$=$O vibrations of each peptide group along the backbone of the peptide that forms a quasi one-dimensional lattice.\cite{Miyazawa:1960uq} These excitons are strongly coupled to the  CO$\cdots$NH hydrogen-bonds that stabilize the helix. This coupling was first introduced by A. S. Davydov which speculated the formation of a soliton able to transfer energy from one side of the $\alpha$-helix to the other.\cite{Davydov:1973qr,Davydov:1985aa,Scott:1982gn,Scott:1992ty} It appeared later that this coupling results into the formation of a vibrational polaron rather than a soliton.\cite{Brown:1989ve,Ivic:1997lq,Pouthier:2003fj,Pouthier:2004fk,Falvo:2005ul,Falvo:2005qf,Falvo:2005ve}\par
Just a few years after the work of Davydov, the infrared (IR) spectroscopy of the molecular crystals acetanilide (ACN)\cite{Careri:1983vn,Careri:1984fd} and N-methylacetamide (NMA)\cite{Fillaux:1981uq,Barthes:1998fk,Herrebout:2001uq} showed some anomalous temperature dependance that was interpreted as a signature of self-trapping. These molecular crystals which consist of quasi-one-dimensional chains of hydrogen-bonded peptide groups resembling the hydrogen-bond network of $\alpha$-helices were then considered as model systems for polypeptides. In ACN, at ambiant temperature, the amide-I band is characterized by a single band located at 1666~\cm, while at lower temperature a second band appears at 1650~\cm. The amide-A band corresponding to the N$-$H stretching vibration is characterized by a main band located at 3295~\cm\ with a series of 9 satelite peaks towards low frequency. These observations show that a strong coupling occurs between the C$=$O and N$-$H vibrations with some low frequency optical phonons. From a theoretical point of view the dynamics of vibrational excitons in molecular crystals and in $\alpha$-helices can be described by the same Holstein Hamiltonian, which was first introduced to describe the dynamics of electrons in molecular crystals. The main difference is that in molecular crystals the vibrational excitons are coupled to optical phonons while in the Davydov Hamiltonian the vibrational excitons are coupled to acoustical phonons.\cite{Eilbeck:1984kx,Alexander:1986yu,Scott:1982gn,Scott:1992ty}\par
Two decades after these observations, Hamm and Edler shed new light on the dynamics of vibrational polarons by performing non-linear IR spectroscopy of ACN and NMA crystals as well as a model $\alpha$-helix.\cite{Edler:2002uq,Edler:2002kc,Edler:2003vg,Hamm:2006uq,Edler:2004qy,Edler:2005gf} This work was reviewed in Ref.~\citenum{Hamm:2009kx}. Developed over the past two-decades, time-resolved nonlinear IR spectroscopy, in particular two-dimensional (2D) IR spectroscopy  have allowed researchers to study the vibrational dynamics of condensed phase systems including peptides, proteins, and water.\cite{Fayer:2013ad,Hamm:2011cr,Khalil:2003ys,Loparo:2006vn,Wong:2013rt,Bloem:2012rr,Middleton:2012hb,Bandaria:2010qo,Ghosh:2014ye,Kim:2009qf} 2D-IR spectroscopy can probe vibrational anharmonic  couplings, vibrational relaxation, population transport, chemical-exchange dynamics and spectral diffusion therefore providing much more information than absorption spectroscopy.\cite{Jansen:2009vf,Zheng:2005jl,Falvo:2008mz,Cho:2008fk,Kim:2009qf} In ACN and NMA, Edler and Hamm used 2D-IR spectroscopy to show that vibrational self-trapping and the formation of vibrational polarons occured in these molecular crystals.\cite{Edler:2002uq,Edler:2002kc,Edler:2003vg,Hamm:2006uq,Hamm:2009kx} They also performed  pump-probe spectroscopy on a model $\alpha$-helix in the N$-$H spectral range.\cite{Edler:2004qy,Edler:2005gf} They show the appearance in the two-exciton spectrum of two bound states. These two bound states were interpreted as the signature of a strong coupling between the vibrational exciton and a set of accoustical phonons in accordance with the original model of Davydov. A similar observation was made a few years later in the spectrum of a model $\beta$-sheet peptide.\cite{Bodis:2009uq}\par
A large number of theoretical studies have been devoted to the Holstein Hamiltonian (electronic or vibrational). Holstein polarons are usually described within two limiting cases that depend on the size of the polaron wavefunction: small and large polarons.\cite{Holstein:1959mz,Holstein:1959kx} For the former the discreteness of the lattice plays a key role while for the later a continuum approximation can be used. Large polarons are often described within the adiabatic limit, i.e. the case when the lattice deformation remains static when the exciton moves along the lattice. In this case, variational approaches\cite{Lu:1991fv,Huynh:2013fk,Sun:2015fk} or mixed quantum-classical simulations\cite{Scott:1992ty,Cruzeiro:2009oz} gives in general good results.\cite{Brown:1989ve,Ivic:1997lq} In contrast, small polarons are usually described within the anti-adiabatic limit which corresponds to a weak exciton coupling.\cite{Tempelaar:2013fj} In this limit the lattice deformation follows the exciton modifying its effective mass. In this case, a canonical transformation allows to switch to the small polaron point of view.\cite{Lang:1963fk,Brown:1989ve,Ivic:1997lq,Pouthier:2003fj,Pouthier:2004fk,Falvo:2005ul,Falvo:2005qf,Falvo:2005ve,Yalouz:2017fk}
It has been shown that for the case of vibrational excitons, because the hopping constant between nearest-neighbor lattice sites is in general small compared to the phonon frequency, the anti-adiabatic limit gives good results.\cite{Brown:1989ve,Ivic:1997lq,Pouthier:2003fj}
The Holstein Hamiltonian has been also solved by a variety of numerical methods that include the Density Matrix Renormalization Group\cite{Jeckelmann:1998oq},  the Multi-Configuration Time-dependant Hartree method,\cite{Tsivlin:2007qf} the Hierarchy Equation Of Motion (HEOM)\cite{Chen:2015uq}, using the two-particle approximation\cite{Philpott:1971aa,Spano:2002dq} or using a direct exact diagonalization.\cite{Hamm:2006uq,Yalouz:2017fk,Yalouz:2017wo} Most theoretical studies focused on the energy transport properties of polarons and on linear spectroscopy, very few are dedicated to predict  non-linear spectroscopy. This is  particularly true for the case of vibrational polarons where to my knowledge the few studies were conducted on pump-probe spectroscopy,\cite{Edler:2004qy,Tsivlin:2006ve,Tsivlin:2006qf,Woutersen:2007fk} and none were conducted on 2D-IR spectroscopy. Note that two recent studies focused on the 2D spectroscopy of electronic excitons in molecular aggregates using a variational approach.\cite{Huynh:2013fk,Sun:2015fk} However, as mentioned earlier this approach is mostly adapted for the case of large polarons and is not adapted for vibrational excitons. Therefore, for vibrational polarons there is a clear lack of theoretical work to predict the non-linear IR response.\par
In this article, the theory of linear and non-linear spectroscopy of vibrational polarons in one-dimensional lattices is presented in order to establish a physical framework to identify the spectral signatures of self-trapping phenomena. The case of both optical and acoustical phonons are considered allowing to cover both Davydov and molecular crystals models. This theoretical work relies on the anti-adiabatic approximation which assumes that the vibrational excitons are slow compared to the phonon bath. Note that in this article, the simple case of a one-dimensional (1D) lattice is investigated keeping the Holstein Hamiltonian as simple as possible in order to set up the framework for the nonlinear response of vibrational polarons and present analytical results. 
In section \ref{sec:theory}, simple expressions are derived for the linear and non-linear optical response of vibrational polarons. These expressions are used in the section \ref{sec:results} for a variety of parameters values in the Holstein model. In section \ref{sec:discussion} where further theoretical derivations are performed, the model results are discussed within the context of experimental observations. Finally, future experiments to probe self-trapping phenomena are suggested in addition to theoretical developments needed in the future as well as conclusions are presented in section \ref{sec:conclusions}.
\section{Theoretical model}
\label{sec:theory}
In this section, the theoretical framework describing vibrational excitons coupled to optical and acoustical phonons is presented within the context of the anti-adiabatic limit. Using this approximation the linear and third-order response is given.
\subsection{Vibrational Holstein Hamiltonian}
A one-dimensional chain of $N$ identical high frequency vibrations coupled to a phonon bath is considered. The system Hamiltonian $\hat{H}$ is written as 
\begin{equation}
\hat{H} = \hat{H}_v + \hat{H}_b + \hat{H}_{vb},
\end{equation}
where $\hat{H}_v$ is the vibrational Hamiltonian, $\hat{H}_b$ the bath Hamiltonian and $\hat{H}_{vb}$ the coupling between the vibrations and the bath. The vibrational Hamiltonian $\hat{H}_v$ is described by an excitonic Hamiltonian written as
\begin{equation}
\hat{H}_v = \sum_{n} \omega_0 b_n^\dagger b_n - A b_n^{\dagger 2} b_n^2 + J \left( b_{n+1}^\dagger + b_{n-1}^\dagger \right) b_n ,
\label{eq:hvibron}
\end{equation}
where $\omega_0$ is the fundamental frequency, $A$ is the anharmonicity, $J$ is the hopping constant and where  $b_n^\dagger$ and $b_n$ are the vibron creation and annihilation operators. In Eq.~(\ref{eq:hvibron}) and in the remaining of this paper, the convention $\hbar=1$ is used. The bath is described by a set of $N$ phonons of frequencies $\Omega_q$ and wavevector $q=2\pi p / N $ with $p=-(N-1)/2,\dots,(N-1)/2$. Using the phonon creation and annihilation operators $a^\dagger_q$ and $a_q$ the bath Hamiltonian is written
\begin{equation}
\hat{H}_b = \sum_{q} \Omega_q (a_q^\dagger a_q + 1/2).
\label{eq:Hb}
\end{equation}
To describe the coupling between the high frequency vibrations and the bath modes it is assumed that each bath mode induces fluctuations of the fundamental frequencies. The coupling hamiltonian is then written as
\begin{equation}
\hat{H}_{vp} = \frac{1}{\sqrt{N}} \sum_{n} \sum_{q} \left( \Delta_{q} e^{-\ii q n} a_{q}^\dagger + \Delta_{q}^{*} e^{\ii q n} a_q \right) b_n^\dagger b_n.
\label{eq:Hvb}
\end{equation}
Note that here each bath modes are coupled to all the vibrations therefore introducing strong bath mediated correlations between different vibrations.
In this article,  two types of phonon models are considered, a model of optical phonons and a model of acoustical phonons. Derivation of the optical and acoustical models are detailed in appendix~\ref{app:phonon}. For the optical phonon model the phonon frequency and coupling are written as
\begin{align}
&\Omega^{\text{opt}}_q = \Omega_{\text{opt}},  \\
&\Delta^{\text{opt}}_q = \Delta_{\text{opt}},
\end{align} 
where $\Omega_{\text{opt}}$ is the frequency of the phonon and $\Delta_{\text{opt}}$ is the coupling strength. The acoustical phonon model is derived from the Davydov Hamiltonian and is given by the following parameters
\begin{align}
&\Omega_q^{\text{ac}} = \Omega_{\text{ac}} \left|\sin q /2 \right|, \\
&\Delta^{\text{ac}}_{q} = -2 \ii \Delta_{{\text{ac}}} \sqrt{|\sin q/2|} \cos q/2,
\end{align}
where $\Omega_{\text{ac}}$ is the cutoff frequency and where $\Delta_{\text{ac}}$ is the coupling strength.
\subsection{Effective Hamiltonian in the anti adiabatic limit}
To partially remove the vibron-bath coupling Hamiltonian, a Lang-Firsov transformation is applied.\cite{Lang:1963fk} A ``full dressing" is considered and the following unitary transformation is introduced
\begin{equation}
\hat{U} = \exp\left( \sum_{n} \hat{X}_n  b_n^\dagger b_n \right),
\label{eq:transfo}
\end{equation}
where the operator $\hat{X}_n$ is defined by
\begin{equation}
\hat{X}_n = \frac{1}{\sqrt{N}}\sum_{q} \left( \frac{\Delta_{q}e^{-\ii q n}}{\Omega_q} a_q^\dagger -  \frac{\Delta_{q}^{*}e^{\ii q n}}{\Omega_q} a_q \right). 
\end{equation}
By using Eq.~(\ref{eq:transfo}), the transformed Hamiltonian $\tilde{H} = U \hat{H} U^\dagger$ is written as
\begin{multline}
\tilde{H} =  \sum_{n} \left(\omega_0 - \epsilon_{0}\right) b_n^\dagger b_n - \left( A + \epsilon_{0} \right)  b_n^{\dagger2} b_n^2 
- 2 \sum_{n < m} \epsilon_{|n-m|} b_n^\dagger  b_n  b_m^\dagger b_m  \\ + \sum_{n} J  \left( \Theta^\dagger_{n+1} b_{n+1}^\dagger + \Theta^\dagger_{n-1} b_{n-1}^\dagger \right)\Theta_{n}b_n  + \hat{H}_b,
\label{eq:hpol}
  \end{multline}
where the dressing operators $\Theta^\dagger_n$  are defined by the transformation of the vibron creation operators $b^\dagger_n$
\begin{equation}
\hat{U} b^\dagger_n \hat{U}^\dagger = b^\dagger_n \Theta^\dagger_n,
\end{equation}
and are written as
\begin{equation}
\Theta^\dagger_n = \exp\left( \hat{X}_n\right).
\end{equation}
The parameters $\epsilon_{n}$ characterize the reorganizational energies of the bath, they are written as
\begin{equation}
\epsilon_{n} = \frac{1}{N}\sum_{q} \frac{\left|\Delta_{q}\right|^2 }{\Omega_q} \cos (n q).
\label{eq:epsn}
 \end{equation}
In the small polaron point of view, the vibrational exciton are dressed by the bath. The remaining coupling between the polaron and the bath now operates through the hopping term which is modulated by bath coherent states. The main advantage of this procedure is that the exciton-phonon coupling has been strongly reduced and a mean field approach can then be used.\cite{Ivic:1997lq}
The final Hamiltonian is written as a sum of three contribution
\begin{equation}
\tilde{H} = \hat{H}_{0} + \hat{H}_b + \Delta\hat{H},
\end{equation}
where $\hat{H}_{0} = \langle \tilde{H} - \hat{H}_b \rangle_b $ is the effective Hamiltonian of the dressed excitons and $\Delta\hat{H} = \tilde{H} - \hat{H}_b - \hat{H}_0$ is the remaining part of the exciton-bath interaction. The symbol $\langle \dots \rangle_b$ stands for the thermal average over the bath degrees of freedom which are assumed to be in equilibrium at temperature $T$. After straightforward calculation, the effective polaron hamiltonian is finally written as
\begin{multline}
\hat{H}_{0} =   \sum_{n} \left(\omega_0 - \epsilon_{0}\right) b_n^\dagger b_n - \left( A + \epsilon_{0}  \right)  b_n^{\dagger2} b_n^2 \\
- 2 \sum_{n< m} \epsilon_{|n-m|} b_n^\dagger  b_n  b_m^\dagger b_m + \sum_{n} J e^{-S(\beta)} \left( b_{n+1}^\dagger + b_{n-1}^\dagger\right) b_n, 
\label{eq:heff}
\end{multline}
where the temperature dependent coupling constant $S(\beta)$ is the nearest-neighbor dressing factor given by
\begin{equation}
S(\beta) = \frac{1}{N}\sum_q  \left| \frac{\Delta_{q}}{\Omega_q} \right|^2 \coth \left( \frac{\beta \Omega_q}{2} \right) \left( 1 - \cos(q) \right),
\end{equation}
where $\beta = 1/k_{\text{B}}T$. In the following, the effect of the remaining coupling $\Delta \hat{H}$ is disregarded and the linear and nonlinear optical responses of polarons will be computed under the effective Hamiltonian given by Eq.~(\ref{eq:heff}). This approximation is relevant in the anti-adiabatic limit where the hopping constant $J$ is small. This approach can be improved by treating the remaining coupling using perturbation theory\cite{Pouthier:2004fk,Pouthier:2013fk,Yalouz:2016aa} which can give very reliable results on a large range of parameters provided that no accidental resonances occur.\cite{Pouthier:2013fk} However, as a first step this work will only consider the effective Hamiltonian and the effect of the remaining coupling will be the subject of a separate study.\par 
Since $\hat{H}_{0}$ commute with the number operator $\hat{N} = \sum_{n} b_n^\dagger b_n$, $\hat{H}_0$ is block diagonal in the eigenvalues of the operator $\hat{N}$, $v=0,1,2,\dots$. This article focus on the third-order nonlinear optical response and therefore only the blocks $v=0,1$ and $v=2$ need to be considered. The one-exciton states block $v=1$  is trivially diagonalized and the eigenstates are expressed by plane-waves
\begin{equation}
\ket{k} = \frac{1}{\sqrt{N}} \sum_{n} e^{\ii k n} b^\dagger_n \left| \varnothing \right\rangle,
\end{equation}
and the eigenvalues are given by
\begin{equation}
\omega_k = \tilde{\omega}_0 + 2 \tilde{J}(\beta) \cos k,
\label{eq:polarondisp}
\end{equation}
where $\tilde{\omega}_0 = \omega_0 - \epsilon_0 $ is the shifted frequency and $\tilde{J}(\beta) = Je^{-S(\beta)}$ is the effective hopping constant.
The two-excitons states block $v=2$ can be simplified by using the periodicity of the lattice. Introducing the following center of mass plane-wave basis\cite{Pouthier:2003fj}
\begin{equation}
\ket{k\  m } =  \frac{1}{\sqrt{N}}  \sum_{n} e^{\ii k \left( n + m /2 \right)} \xi_m b_n^\dagger b_{n+m}^\dagger \left| \varnothing \right\rangle,
\end{equation}
where $m=0,\dots,(N-1)/2$ is the distance between the two-exciton and where $\xi_m$ is defined by
\begin{equation}
\xi_m =  \begin{cases} 0  & \mbox{if } m<0,  \\  1/\sqrt{2} & \mbox{if } m=0, \\ 1 & \mbox{if } m>0 . \end{cases}
\end{equation}
Using this basis-set, one can show that the Hamiltonian is block diagonal in the wave vector $k$. The $k$-th block can be deduced from the equations\cite{Pouthier:2003fj}
\begin{align}
\hat{H}_0\ket{k\  0} &= \left( 2\tilde{\omega}_0 - 2A - 2\epsilon_0 \right) \ket{k\  0 } + \sqrt{2} \tilde{J}_k  \ket{k\  1 },   \\ 
\hat{H}_0\ket{k\  1} &=  \left(2\tilde{\omega}_0 - 2\epsilon_1 \right)  \ket{k\  1 } +  \sqrt{2} \tilde{J}_k  \ket{k\  0 } + \tilde{J}_k  \ket{k\  2 }, \\
\hat{H}_0\ket{k\  m} &=  \left(2\tilde{\omega}_0 - 2\epsilon_m \right)  \ket{k\  m } + \tilde{J}_k  \ket{k\  m-1 } + \tilde{J}_k  \ket{k\  m+1 }, & \mbox{if } m>1 ,
\end{align}
with $\tilde{J}_k = 2\tilde{J}\cos(k/2)$. Each block $k$ of $\hat{H}_0$ can then be easily diagonalized numerically giving a set of eigenvalues $\omega_{k\sigma}$ and eigenvectors $\psi_{k\sigma}(m)$ where $\sigma=0,\dots,(N-1)/2$ labels the different eigenvalues.
\subsection{Linear optical response}
The coupling of the vibrations to the optical field $E(\textbf{r},t)$  is given by
\begin{equation}
\hat{H}_{\text{int}} = -E(\textbf{r},t) \hat{V},
\end{equation}
where $\hat{V}$ is the dipole operator expressed for a set of identical molecules as a function of the projection of the transition dipole moments $\mu$ on the electric field as
\begin{equation}
\hat{V} = \sum_{n} \mu \left( b_n + b_n^\dagger \right).
\end{equation}
The linear optical response is given by the response function written as
\begin{equation}
R^{(1)}(t) = \ii\Theta(t) \left\langle \left[ \hat{V}(t), \hat{V}\right] \right\rangle = \ii \Theta(t) \left( J(t) - J^*(t) \right) ,
\label{eq:r1t}
\end{equation}
where $\Theta(t)$ is the Heaviside function, $V(t) = e^{\ii \hat{H} t} \hat{V}e^{-\ii \hat{H}t}$ is the time evolution of the dipole operator in the Heisenberg picture and where $\langle \dots \rangle$ is the thermal average over all degrees of freedom. The function $J(t)$ can be expressed as a function of the total density matrix at equilibrium $\hat{\rho} = \exp(-\beta \hat{H})/Z(\beta)$ as
\begin{equation}
J(t) = \left \langle \hat{V}(t) \hat{V} \right\rangle = \tr \left[ \hat{\rho} e^{\ii \hat{H} t} \hat{V} e^{-\ii \hat{H} t}  \hat{V} \right].
\end{equation}
By introducing the Lang-Firsov unitary transformation $\hat{U}$ in the correlation function, neglecting the remaining coupling $\Delta \hat{H}$, assuming that the harmonic frequencies of the vibrations are much higher than the temperature and using the rotating wave approximation, the function $J(t)$ is now written
\begin{equation}
J(t) = \sum_{n,m} \mu^2 \left\langle \varnothing \right | b_n e^{-\ii \hat{H}_{0}t}  b^\dagger_m \left| \varnothing \right\rangle C_{|n-m|}(t)
\end{equation}
where $C_{|n-m|}(t)$ is the bath correlation function given by
\begin{equation}
C_{|n-m|}(t) = \left\langle \theta_n(t)  \theta_m^\dagger \right\rangle_b = \exp\left(-g_{|n-m|}(t)\right),
\end{equation}
where the linebroadening function $g_{n}(t)$ is given by
\begin{equation}
g_{n}(t) = \frac{1}{N}\sum_{q}  \left| \frac{\Delta_{q}}{\Omega_q} \right|^2  \left\{\coth\left( \frac{\beta \Omega_q}{ 2} \right) \left( 1 -\cos\left( \Omega_q t - q n \right)  \right) +  \ii \sin\left( \Omega_qt -qn\right) \right\}.
\label{eq:gnt}
\end{equation}
Note that the linebroadening function for $n=1$ at $t=0$ is simply the nearest-neighbor dressing factor $g_1(0) = S(\beta)$.  After straightforward calculation, the optical response is written
\begin{equation}
J(t) =  \mu^2 \sum_k e^{-\ii \omega_k t} C_k (t),
\label{eq:Jt}
\end{equation}
where $C_k(t)$ is the spatial Fourier transform of the bath correlation function
\begin{equation}
C_k(t) = \sum_n e^{\ii k n } \exp\left(-g_{n}(t)\right).
\label{eq:Ck}
\end{equation}
Finally, the absorption spectrum $\alpha(\omega)$ is then directly proportional to the Fourier transform of the response fonction $R^{(1)}(t)$ given by Eq.~(\ref{eq:r1t}).
For a periodic and isolated system, assuming that the laser wavelength is larger than the system's typical size, only the excitation energy corresponding to a vanishing wavevector $k\rightarrow0$ should contribute to the linear optical response. Here, because of the coupling to the phonon bath, all modes $k$ contribute to the optical response with different weights corresponding to the spatial Fourier transform of the bath correlation function. Note that the expression for the linebroadening function $g_n(t)$ is very close to the expression for the linebroadening function of a single isolated transition coupled to a harmonic bath.\cite{Duke:1965cr} In Eq.~(\ref{eq:gnt}), the dephasing of the vibrations takes into account the delocalized nature of the phonons and the  correlation induced by the bath. In addition, the Stokes shift which is usually included in the definition of the linebroadening function\cite{Mukamel:1995fk} is not present in Eq.~(\ref{eq:gnt}). This Stokes shift is in fact included directly in the definition of the polaron Hamiltonian via the reorganizational energies $\epsilon_n$ defined in Eq.~(\ref{eq:epsn}).
\subsection{Third-order nonlinear optical response}
The third-order response function is given by
\begin{equation}
R^{(3)}(t_1,t_2,t_3) = \ii^3  \Theta(t_1) \Theta(t_2) \Theta(t_3) \left\langle \left[ \left[ \left[ \hat{V}(t_1+t_2+t_3), \hat{V}(t_1+t_2)\right], \hat{V}(t_1) \right], \hat{V}(0) \right] \right\rangle.
\end{equation}
The three nested commumators yield eight Liouville space pathways.\cite{Mukamel:1995fk,Abramavicius:2009fk} Each nonlinear technique is based on a specific phase matching condition which selects a subgroup of pathways. For simplification, only the expressions for the signal corresponding to the direction $\textbf{k}_{\text{I}} = -\textbf{k}_1 + \textbf{k}_2 + \textbf{k}_3$ are presented. The expression for the direction $\textbf{k}_{\text{II}} = \textbf{k}_1 - \textbf{k}_2 + \textbf{k}_3$  is given in appendix~\ref{app:signals}.
The total response function for $\textbf{k}_{\text{I}}$ is given as a sum of three contributions
\begin{equation}
R_{\textbf{k}_{\text{I}}}(t_1,t_2,t_3) = R_1 + R_2 - R_3,
\end{equation}
where $R_1$ is the ground state bleaching (GSB), $R_2$ is the excited state emission (ESE) and $R_3$ is the excited state absorption (ESA). Using the same approach as for the linear response function each contributions to the $\textbf{k}_{\text{I}}$ signal can be written as
\begin{align}
R_1(t_1,t_2,t_3) &= \frac{\mu^4}{N} \sum_{k_1 k_2} e^{ \ii \omega_{k_1} t_1-\ii \omega_{k_2} t_3 } C^{(1)}_{k_1 k_2}(t_1,t_2,t_3), \\
R_2(t_1,t_2,t_3) &= \frac{\mu^4}{N} \sum_{k_1 k_2} e^{ \ii \omega_{k_1} (t_1+t_2)-\ii \omega_{k_2} (t_3+t_2) } C^{(2)}_{k_1 k_2}(t_1,t_2,t_3), \\
R_3(t_1,t_2,t_3) &= \frac{\mu^4}{N^2}\sum_{k_1 k_2 k_3 \sigma} e^{ \ii \omega_{k_1} (t_1+t_2+t_3) -\ii \omega_{k_2} t_2 -\ii \omega_{k_3 \sigma} t_3 } C^{(3)}_{k_1 k_2 k_3}(t_1,t_2,t_3) A_{k_1 k_3 \sigma} A_{k_2 k_3 \sigma},
\end{align}
with the functions $C^{(i)}(t_1,t_2,t_3)$ defined by
\begin{align}
C^{(1)}_{k_1 k_2} (t_1,t_2,t_3) &= \sum_{m_1 m_2 m_3} e^{-\ii k_1 m_1 + \ii k_2 m_3} e^{-g^{(1)}_{m_1 m_2 m_3}(t_1,t_2,t_3) }, \label{eq:Ck1}\\
C^{(2)}_{k_1 k_2} (t_1,t_2,t_3) &=  \sum_{m_1 m_2 m_3} e^{-\ii k_1 (m_1+m_2) + \ii k_2 (m_2+ m_3) } e^{-g^{(2)}_{m_1 m_2 m_3}(t_1,t_2,t_3) },\label{eq:Ck2} \\
C^{(3)}_{k_1 k_2 k_3} (t_1,t_2,t_3) &= \sum_{m_1 m_2 m_3} e^{-\ii k_1 (m_1+m_2+m_3) + \ii k_2 m_2 + \ii k_3 m_3} e^{-g^{(3)}_{m_1 m_2 m_3}(t_1,t_2,t_3) },\label{eq:Ck3}
\end{align}
and where the linebroadening functions are given by
\begin{align}
&g^{(1)}_{m_1m_2m_3}(t_1,t_2,t_3) = g^*_{m_1}(t_1) -g^*_{m_2}(t_2) +  g_{m_3}(t_3) \nonumber \\ 
     &+ g^*_{m_1+m_2}(t_1+t_2) +  g^*_{m_2+m_3}(t_2+t_3)  -g^*_{m_1+m_2+m_3}(t_1+t_2+t_3), \\
&g^{(2)}_{m_1m_2m_3}(t_1,t_2,t_3) = g^*_{m_1}(t_1) -g_{m_2}(t_2) +  g^*_{m_3}(t_3) \nonumber \\ 
     &+ g^*_{m_1+m_2}(t_1+t_2) +  g_{m_2+m_3}(t_2+t_3)  -g^*_{m_1+m_2+m_3}(t_1+t_2+t_3), \\
&g^{(3)}_{m_1m_2m_3}(t_1,t_2,t_3) = g^*_{m_1}(t_1) -g_{m_2}(t_2) +   g_{m_3}(t_3) \nonumber \\ 
     &+ g^*_{m_1+m_2}(t_1+t_2) +  g_{m_2+m_3}(t_2+t_3)  -g^*_{m_1+m_2+m_3}(t_1+t_2+t_3).
\end{align}
The tensor $A_{k k' \sigma}$ is expressed as a function of the two-excitons wave function
\begin{equation}
A_{k k' \sigma} = 2 \sum_m \Psi_{k'\sigma}(m) \xi_m \cos\left( \left( k'/2 - k \right) m \right).
\end{equation}
Note that the discrete spatial Fourier transform in Eqs.~(\ref{eq:Ck}), (\ref{eq:Ck1}), (\ref{eq:Ck2}) and (\ref{eq:Ck3}) can be easily computed numerically using the 1D, 2D and 3D Fast Fourier Transform (FFT) algorithm.\cite{Frigo:2005uq}
\section{Results}
\label{sec:results}
In this section, the previous formalism is applied to compute the linear and nonlinear spectroscopy of vibrational polarons in a 1D lattice. The parameters range used here corresponds to the amide-I vibration in $\alpha$-helix polypeptides and molecular crystals such as ACN or NMA often modeled as quasi-one-dimensional chains. The intramolecular anharmonicity value is fixed to $A=8~\cm$.\cite{Hamm:1998du} The hopping constant ranges between -10~\cm~to 10~\cm.~\cite{Hamm:2006uq} For the optical phonon model the optical phonon frequency is fixed to $\Omega_{\text{opt}}=~50~\cm$ corresponding to the crystalline acetanilide (ACN) optical frequency.~\cite{Hamm:2006uq} For the acoustical phonon model, the cutoff frequency is fixed to $\Omega_{\text{ac}}=100~\cm$ corresponding to the $\alpha$-helices cutoff frequency.\cite{Scott:1982gn} The coupling between the vibration and the phonons will take a typical value of $\Delta_{\text{opt}}=\Delta_{\text{ac}}=25~\cm$.~\cite{Hamm:2006uq}
A phenomenological life-time of $T_1=1.5$ ps was added to the calculation of the linear and non-linear spectra. This value is chosen close to the relaxation time of 1.2 ps measured for amide-I vibration in peptides.\cite{Hamm:1998du} All numerical calculations were performed using a number of sites of $N=51$. This number was found to be large enough to obtain results close to  an infinite system.  For the case of the acoustical phonon model, to avoid spurious effects due to the finite size used in the numerical calculations, the sum over the phonon wavevector $q$ in Eq.~(\ref{eq:gnt}) is performed using a larger number of phonon modes $N_{\text{ph}} = 5001$. This number is chosen so that no recursion is observed in the behavior of the linebroadening function $g_n(t)$ given by Eq.~(\ref{eq:gnt}). Finally the harmonic frequency is set to the value $\omega_0=0$ without any loses of generality.
In the following subsections, the influence of the structure of the bath on the linear and non-linear vibrational responses is investigated by using  optical and acoustical phonon models. 
\subsection{Optical phonon model}
\begin{figure}
\includegraphics{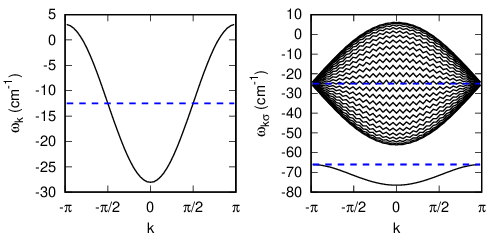}
\caption{One-polaron (left panel) and two-polarons (right panel) energy spectra for the optical phonon model as a function of the wave vector $k$ for  $\Omega_{\text{opt}}=~50~\cm$, $\Delta_{\text{opt}}=25~\cm$, $T=0$ K and for $J=-10~\cm$ (black solid lines) and $J=0~\cm$ (blue dashed lines).}
\label{fig:energy_opt}
\end{figure}
First, the case of the optical phonon model is considered. The one-polaron energy spectrum which controls the behavior of the linear absorption spectrum is depicted on the left panel of Fig.~\ref{fig:energy_opt}. As seen in Eq.~(\ref{eq:polarondisp}), the one-polaron eigenfrequencies are centered around the shifted frequency $\tilde{\omega}_0$ with a width of $4\tilde{J}(\beta)$. For $\Omega_{\text{opt}}=~50~\cm$, $\Delta_{\text{opt}}=25~\cm$, $T=0$ K and $J=-10~\cm$, the shifted frequency takes the value $\tilde{\omega}_0 = -12.5~\cm$ and the total dispersion is $4\tilde{J} = 31.2~\cm$. The two-polarons energy spectrum which controls the behavior of the non-linear optical response is reported on the right panel of Fig.~\ref{fig:energy_opt} for the same set of parameters. The energy spectrum shows one continuum band characterizing the two-polaron free states with a total bandwidth of $8\tilde{J} = 62.4~\cm$ and an isolated band corresponding to the two-polaron bound states which is determined by the anharmonicity.\cite{Kimball:1981fk}

\begin{figure*}
\includegraphics{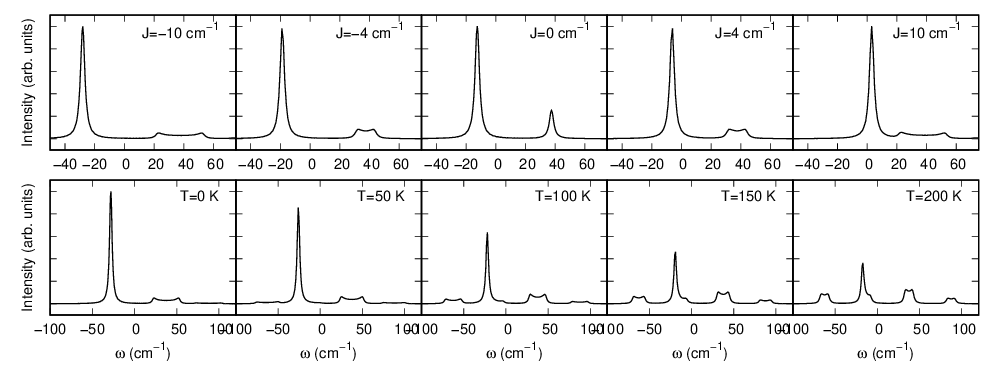}
\caption{Upper panel: Linear absorption spectrum of the optical phonon model for $\Omega_{\text{opt}}=~50~\cm$,  $\Delta_{\text{opt}}=25~\cm$ and $T=0$ K  as a function of the value of the hopping constant $J$. Lower panel: Linear absorption spectrum for $\Omega_{\text{opt}}=~50~\cm$,  $\Delta_{\text{opt}}=25~\cm$ and $J=-10~\cm$ as a function of the temperature $T$.}
\label{fig:linear_opt}
\end{figure*}
The upper panel of Fig.~\ref{fig:linear_opt} shows the dependence of the linear absorption spectrum for the optical phonon model as a function of the hopping constant $J$. For $J=-10$~\cm, the spectrum exhibits a sharp and strong zero-phonon line (ZPL) located at $\omega=-28.1~\cm$ corresponding the the position of the lowest polaron energy $\omega_{k=0}=\tilde{\omega}_0-2\tilde{J}$. A broad second band is also present in the absorption spectrum. It corresponds to the one-phonon excitation band (0-1 transition in the Franck-Condon (FC) picture) and is shifted by $\Omega_{\text{opt}}$ with respect to the ZPL. This broad band exhibits a double peak shape and has a bandwidth of 31~\cm\ corresponding to the total polaron dispersion $4\tilde{J}$.  As decreasing the value of $|J|$ to 0 the ZPL shifts towards $\omega=-12.5~\cm$ while the second band does not shift but its bandwidth reduces significantly. Upon increasing the value of $J$ to 10~\cm, the ZPL continues to shift to $\omega=0~\cm$ and the one-phonon band bandwidth increases again to recover the bandwidth for $J=-10$~\cm. This behavior of the linear absorption spectrum is almost identical to the numerical calculation performed by Hamm and Edler based on a direct diagonalization of the Holstein Hamiltonian.~\cite{Hamm:2006uq} The main difference between this result and the result of Ref.~\citenum{Hamm:2006uq} is seen in the one-phonon band which is symmetric in this calculation but is stronger on the lower energy side of the band in Ref.~\citenum{Hamm:2006uq}. This difference can be explained by the remaining coupling term $\Delta\hat{H}$ which was neglected here and which introduce a residual coupling between the ZPL and the one-phonon band. However, this result captures the main features observed in Ref.~\citenum{Hamm:2006uq}. The lower panel of Fig.~\ref{fig:linear_opt} shows the temperature dependance of the linear absorption spectrum for the optical phonon model. Upon increasing the temperature, the linear absorption spectrum exhibits new bands on the blue side of the spectrum corresponding to $n$th phonon bands as well as hot bands on the red side of the spectrum. Also by increasing the temperture, the ZPL decreases sharply. All $n$th phonon bands and hot-bands exhibit the same broad double peak shape with a bandwidth which corresponds to the polaron dispersion $4\tilde{J}$ and decreases with temperature. The ZPL shape is also modified as the temperature is increased.\par
\begin{figure}
\includegraphics{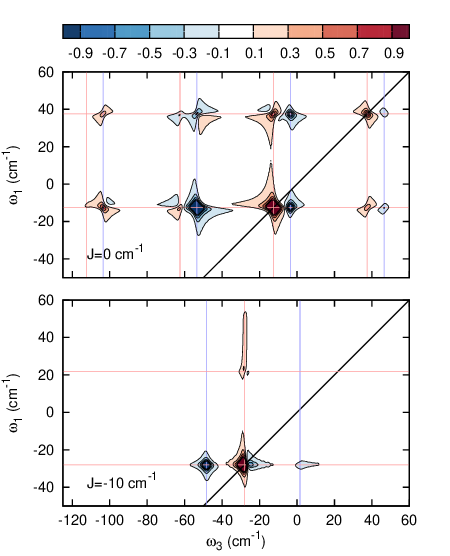}
\caption{2D spectrum for the optical phonon model for $\Omega_{\text{opt}}=~50~\cm$,  $\Delta_{\text{opt}}=25~\cm$ and $T=0$ K and for $J=0$ and $J=-10$~\cm. Vertical red lines and horizontal red lines mark the position of the ZPL and the $n$th-phonon bands. The blue vertical lines correspond to the ZPL and $n$th-phonon lines shifted by the anharmonicity}
\label{fig:2d_opt}
\end{figure}
Fig.~\ref{fig:2d_opt} shows the absorptive part of the 2D spectrum (sum of the rephasing $\textbf{k}_{\text{I}}$ and non-rephasing $\textbf{k}_{\text{II}}$ signals\cite{Hamm:2011cr}) for $J=0~\cm$ and $J=-10~\cm$. For $J=0~\cm$ the 2D spectrum shows multiple negative and positive peaks. Horizontal red lines have been added to mark the position of the ZPL and the one-phonon band in the absorption spectrum. The vertical red lines mark the position of the ZPL, the one-phonon band as well as the one-phonon hot band (1-0 transition in the FC picture). The one-phonon hot band  originates from the excited state emission. The strongest negative peak corresponds to the anharmonically shifted ZPL originating from the excited state absorption. The position of this transition as well as the shifted one-phonon and two-phonons bands and the shifted hot-band are marked by vertical blue lines. Note that the shifted hot-bands appear as positive peaks and not negative peaks even though they originate from the ESA contribution. Analysis of the response functions shows that the vibrational overlaps corresponding to these peaks are negative therefore changing the sign of the peaks.\par
For $J=-10~\cm$ the 2D spectrum is strongly modified as most bands disappear. Only the ZPL and the one-phonon band remain. As in the linear absorption spectrum, the one-phonon band is broad with a bandwidth corresponding to the polaron dispersion. However, this can only being seen along the $\omega_1$ axis of the 2D spectrum. An additional peak appears on the blue side of the ZPL along the $\omega_3$ axis, this peak originates from the exciton-exciton scattering.\cite{Abramavicius:2013kx} A similar peak is also visible on the side of the one-phonon band.

\subsection{Acoustical phonons}
\begin{figure}
\includegraphics{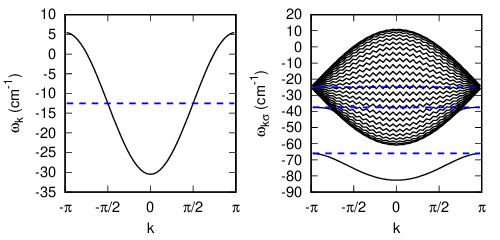}
\caption{One-polaron (left panel) and two-polarons (right panel) energy spectra for the acoustical phonon model as a function of the wavevector $k$ for  $\Omega_{\text{ac}}=~100~\cm$,  $\Delta_{\text{ac}}=25~\cm$ , $T=0$ K and for $J=-10~\cm$ (black solid lines) and $J=0~\cm$ (blue dashed lines).}
\label{fig:energy_ac}
\end{figure}
\begin{figure*}
\includegraphics{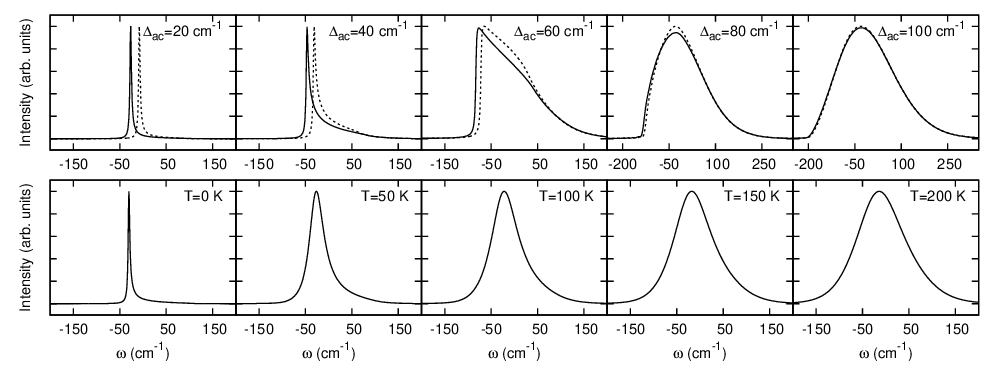}
\caption{Upper panel: Linear absorption spectrum of the acoustical phonon model for $\Omega_{\text{ac}}=~100~\cm$,  $\Delta_{\text{ac}}=25~\cm$ and $T=0$ K  as a function of the value of the phonon coupling $\Delta_{\text{ac}}$ and for $J=0$ (dashed line) and $J=-10~\cm$ (full line).  Lower panel: Linear absorption spectrum for $\Omega_{\text{ac}}=~100~\cm$,  $\Delta_{\text{ac}}=25~\cm$ and $J=-10~\cm$ as a function of the temperature $T$. The spectra are normalized to the maximum of the peak.}
\label{fig:linear_ac}
\end{figure*}
Next, the case of the acoustical phonon model is considered. The one-polaron and two-polarons energy spectra are depicted in Fig.~\ref{fig:energy_ac} for the parameters $\Omega_{\text{ac}}=~100~\cm$,  $\Delta_{\text{ac}}=25~\cm$ , $T=0$ K and for $J=-10~\cm$ and $J=0~\cm$. The main difference in the two-polarons energy spectrum with respect to the optical phonon model can be seen by the appearance of a second bound state. This bound state is present for all wavevector $k$ for $J=0$ but is only present near $k=\pi/2$ for $J=-10~\cm$. The presence of these two bound states are related to the occurence of two type of anharmonicities of local and non-local nature due to the coupling with the phonon bath. Detailed studies on the nature of these bound states have been performed.\cite{Pouthier:2003fj,Falvo:2006ly} For example, a phase diagram for the appearance of the bound states as a function of the anharmonicity, coupling to the bath and hopping constant, have been drawn using decimation methods.\cite{Pouthier:2003fj} The presence of two bound states have been suggested to occur for the N$-$H vibrations in $\alpha$-helix and $\beta$-sheet peptides trough the appearance of two excited state absorption peaks in their pump-probe spectrum.\cite{Edler:2004qy,Edler:2005gf,Bodis:2009uq}\par
The upper panel of Fig.~\ref{fig:linear_ac} shows the dependence of the linear absorption spectrum for the acoustical phonon model as a function of the coupling constant $\Delta_{\text{ac}}$ at $T=0$ K and for a hopping constant $J=-10~\cm$ and $J=0~\cm$. In Fig.~\ref{fig:linear_ac}, the spectra were normalized with respect to the maximum of the band to highlight the increase in bandwidth as a function of the coupling.  For small values of the coupling the linear absorption is characterized by  a single asymmetric band. Upon increasing the coupling constant the absorption band bandwidth increases keeping an asymmetric shape. Only at very large value of the coupling the shape of the band tends to be more symmetric. The full width at half maximum (FWHM) of the spectrum increases from $4~\cm$ for $\Delta_{\text{ac}}=20~\cm$ to $215~\cm$ for $\Delta_{\text{ac}}=100~\cm$. For small values of $\Delta_{\text{ac}}$, the effect of the hopping constant $J$ is a simple shift of the band by $2\tilde{J}$. Upon increasing the value of the coupling this shift decreases. The lower panel of Fig.~\ref{fig:linear_ac} shows the temperature dependance of the linear absorption spectrum for the acoustical phonon model. Upon increasing the temperature, the bandwidth of the absorption band increases rapidly and its shape become more symmetric with a typical Lorentzian shape. The bandwidth (full width at half maximum) increases from $5~\cm$ for $T=0$ K to $120~\cm$ for $T=200$ K.\par
\begin{figure}
\includegraphics{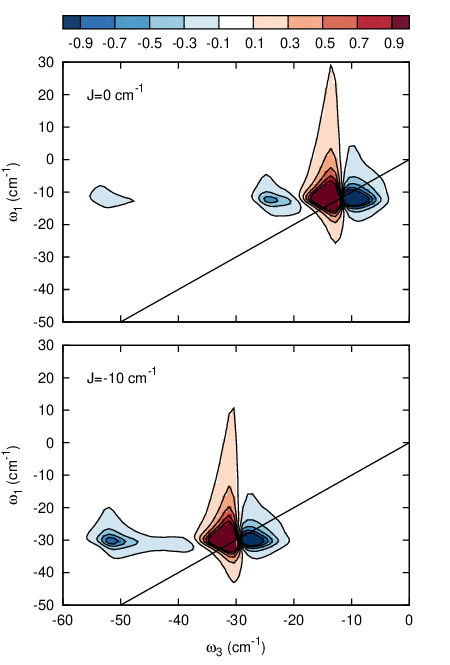}
\caption{2D spectrum for the acoustical phonon model for $\Omega_{\text{ac}}=~100~\cm$,  $\Delta_{\text{ac}}=25~\cm$ and $T=0$ K and for $J=0$ and $J=-10$~\cm.}
\label{fig:2d_ac_2}
\end{figure}
Next, the nonlinear optical response in the low temperature regime is discussed. Fig.~\ref{fig:2d_ac_2} shows the absorptive part of the 2D spectrum for $\Omega_{\text{ac}}=~100~\cm$,  $\Delta_{\text{ac}}=25~\cm$, $T=0$ K and for $J=0$ and $J=-10$~\cm. For $J=0$ the 2D spectrum is characterized by a pair of negative-positive peaks located on the diagonal of the spectrum originating from the interference of the ESA and the GSB and ESE pathways and two negative peaks red shifted along the $\omega_3$ axis. These two peaks are the signature of the two bound states visible in the two-polarons spectrum. For $J=-10~\cm$, the pair of negative-positive peak is red shifted by $2\tilde{J}$ and only one negative peak is present. This results from the fact that only one bound state is present for the wavevector $k=0$.\par
\begin{figure}
\includegraphics{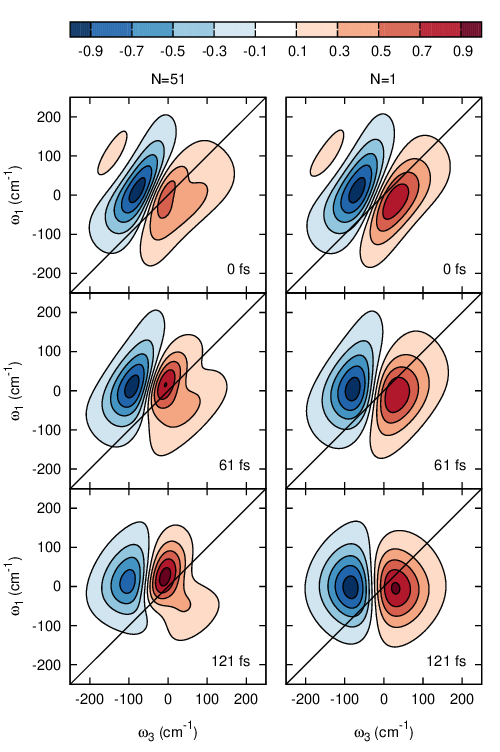}
\caption{2D spectra for the acoustical phonon model for $\Omega_{\text{ac}}=~100~\cm$,  $\Delta_{\text{ac}}=25~\cm$ and $T=300$ K, $J=-10$~\cm\  as a function of the delay time $t_2$, for a 1D chain (left panel) and for a single isolated site $N=1$.}
\label{fig:2d_ac_1}
\end{figure}
At high temperature, the 2D spectrum is strongly modified. The left panel of Fig.~\ref{fig:2d_ac_1} shows the 2D spectrum for $\Omega_{\text{ac}}=~100~\cm$,  $\Delta_{\text{ac}}=25~\cm$, $T=300$ K and for $J=0$ and $J=-10$~\cm\ and for different time delay $t_2$. For $t_2=0$ the 2D spectrum is characterized by a pair of negative-positive  peaks elongated along the diagonal. The width along the diagonal is much larger than the width of the low temperature 2D spectrum and corresponds essentially to the linear absorption bandwidth. The shape of the 2D spectrum is tightly connected to the separation of homogeneous and inhomogeneous broadening. As the waiting time $t_2$ increases, the shape of the two peaks is strongly modified toward a more circular shape. This is a well known phenomenon which has been observed in many molecular systems and is a signature of the spectral diffusion due to fluctuations of the frequencies.\cite{Ishikawa:2007rm,Fecko:2005pd,Falvo:2015ty} In particular the shape of the peaks can be directly related to the frequency-frequency correlation function (FFCF) through metrics computed from 2D lineshape such as the center line slope (CLS).\cite{Kwak:2007fk,Kwak:2008uq,Falvo:2016fk} To understand the effect of the bath-induced correlations, the 2D spectrum for the same time delays $t_2$ and for a single site $N=1$, but still assuming a infinite phonon model, is computed. This would correspond, for example, to the spectroscopy of an impurity.\cite{Duke:1965cr} For a single site, coupling to the bath induces a similar elongated shape. But as $t_2$ increases the shape of the peaks is more rapidly circular. This is a signature of a faster decay of the FFCF. 
\section{Interpretation and discussion}
\label{sec:discussion}
In the previous section, the numerical results have shown that the nature of the bath can strongly modify the linear and non-linear optical lineshape. Specific approximations and analytical expressions can be developed to fully understand the relation between the spectroscopic signature and the processes involved. In this section I will also discuss the physical meaning and the experimental implications of these results.
\subsection{Optical phonons}
For the optical phonon model, the bath coupling and optical frequency are independent of the wavevector $q$. In this case, very simple expressions can be derived for the linear absorption. Introducing the Huang-Rhys coupling constant $S_{\text{opt}} = \Delta_{\text{opt}}^2/\Omega_{\text{opt}}^2$, the Stokes shift and the dressing factor are then expressed as
\begin{align}
&\epsilon_n = \frac{\Delta_{\text{opt}}^2}{\Omega_{\text{opt}}} \delta_{n,0}= \Omega_{\text{opt}}S_{\text{opt}}\delta_{n,0},\\
&S(\beta) = S_{\text{opt}} \coth \left( \frac{\beta \Omega_{\text{opt}}}{2}\right).
\end{align}
The linebroadening functions are written as
\begin{equation}
g_n(t) =  S_{\text{opt}} \left\{  \coth \left( \frac{\beta \Omega_{\text{opt}}}{2} \right)\left( 1 - \delta_{n,0} \cos\Omega_{\text{opt}}t \right) + \ii \delta_{n,0} \sin\Omega_{\text{opt}}t  \right\}.
\end{equation}
After straightforward calculations, the linear response function can be written
\begin{equation}
J(t) = N \mu^2 e^{-S(\beta)} \left[ e^{-\ii \omega_{k=0} t} +  \tilde{\rho}(t) \sum_{n=-\infty}^\infty  M_n(\beta) e^{-\ii n \Omega_{\text{opt}} t} \right],
\label{eq:Jtopt}
\end{equation}
where $\tilde \rho(t)= \frac{1}{N} \sum_k e^{-\ii \omega_k t}$ is the Fourier transform of the polaron density of states and where the constants $M_n(\beta)$ are defined by 
\begin{equation}
M_n(\beta) = I_n \left(S_0/\sinh( \beta \Omega_{\text{opt}}/2)\right) e^{n\Omega_{\text{opt}}\beta/2} - \delta_{n,0},
\end{equation}
where $I_n(x)$ are  the modified Bessel functions of the first kind.\cite{Abramowitz:1972qf} The linear absorption $\alpha(\omega)$ is directly proportional to the Fourier transform of the function $J(t)$ and is written
 \begin{equation}
 \alpha(\omega) =  N \mu^2 e^{-S(\beta)} \left( \delta\left(\omega-\omega_{k=0}\right) + \sum_{n=-\infty}^\infty  M_n(\beta)  \rho\left(\omega-n\Omega_{\text{opt}}\right) \right).
 \label{eq:alphaopt}
 \end{equation}
 Therefore  the absorption spectrum is given by the sum of a delta-like ZPL and a series of peaks corresponding to the Franck-Condon vibrational progression. For $T=0$ K, the ZPL is not broadened by the bath  while the shape  of the other bands is  given by the polaron density of states $\rho(\omega) = \frac{1}{N} \sum_k \delta(\omega-\omega_k)$. From Eq.~(\ref{eq:polarondisp}), it is easy to deduce  the polaron density of states in the limit $N\rightarrow\infty$, it is written
 \begin{equation}
 \rho(\omega) = \frac{1}{\pi \sqrt{4\tilde{J}^2 - \left( \omega - \tilde{\omega}_0 \right)^2}} \quad \text{if} \quad | \omega - \tilde{\omega}_0|  < 2\tilde{J},
 \label{eq:rhopol}
 \end{equation}
which diverge when $\omega = \tilde{\omega}_0 \pm 2\tilde{J}$ resulting in a double peak shape as observed in Fig.~\ref{fig:linear_opt}. The expressions for the linear response obtained here are very similar to the expressions given in Ref.~\citenum{Cevizovic:2009hi} which uses a similar derivation. The main difference appears for the ZPL  which in Ref.~\citenum{Cevizovic:2009hi} is broadened by the phonons even at $T=0$ K while in our case it is not since $M_{0}(\infty) = 0$. Note that in Ref.~\citenum{Cevizovic:2009hi}, the regime explored corresponds to a strong Huang-Rhys coupling constant for which the polaron bandwidth vanishes. Eqs.~(\ref{eq:alphaopt}) and (\ref{eq:rhopol}) fully capture the behavior of the linear absorption in Fig.~\ref{fig:linear_opt} and in Ref.~\citenum{Hamm:2006uq}. Therefore, this work gives a theoretical basis to understand the nature of the peaks observed in the linear absorption.\par
Next, the previous results are used to understand the physical meaning of the experimental measurements of ACN. In ACN, the experimental linear absorption spectrum shows at low temperature one peak located at $1666~\cm$ and a second band at $1650~\cm$. The band at $1666~\cm$ shows some marked features with three subbands which disappear at high temperature and the band at $1650~\cm$  decreases rapidly as a function of temperature (see for example Refs.\citenum{Careri:1984fd}, \citenum{Eilbeck:1984kx}, and \citenum{Hamm:2009kx}). This behavior is very similar to the result presented in the present study using the vibrational Holstein model. As one can expects from the Franck-Condon picture the intensity of the low frequency band which corresponds to the ZPL decreases sharply as a function of temperature while the intensity of the one-phonon band increases.  In addition, the shape of the one-phonon band which corresponds to the polaron density of states decreases as a function of  temperature due to an increase of the dressing factor (Eq.~\ref{eq:rhopol}). This is consistent with the disappearance of the substructure observed in the $1666~\cm$ band of ACN. Note that in the experiment, the frequency difference between the two peaks is much smaller than the frequency difference in the 1D Hostein model. This difference could originate from the fact that we did not include the 3D structure of the crystal and additional dipole-dipole couplings need to be included.\cite{Hamm:2006uq,Hamm:2009kx} Similarly, the experimental 2D-IR spectrum of ACN have some marked similarities with the 2D-IR septrum predicted by the vibrational Holstein model (see for example Refs.~\citenum{Edler:2003vg} and~\citenum{Hamm:2009kx}). It shows clearly a pair of negative/positive peaks located at $1650~\cm$ which would correspond to the ZPL with an aditionnal side peak which we have interpreted as a result of exciton-exciton scattering. In addition there is no peaks on the diagonal corresponding to the $1666~\cm$ band (one-phonon band in the Holstein model).

\subsection{Acoustical phonons}
For the acoustical phonon model, introducing the coupling constant $S_{\text{ac}} = 2\Delta_{\text{ac}}^2 / \Omega_c^2$, the dressing factor and the Stokes shifts are given by 
\begin{align}
&S(\beta) = \frac{8S_{\text{ac}}}{\pi} \int_0^{1} \ud x \  x \sqrt{1-x^2 } \coth\left(\beta\Omega_c x / 2\right), \\
&\epsilon_{n} = \Omega_c S_{\text{ac}} \left( \delta_{n,0} + \frac{1}{2}\delta_{n,1} + \frac{1}{2}\delta_{n,-1}  \right),
\label{eq:anharcouplings}
\end{align}
giving rise to two types of couplings: a local anharmonic coupling and a nearest-neighbor anharmonic coupling. The effect of these two types of anharmonic couplings on the two-exciton states and the transport properties has been studies in details.\cite{Pouthier:2003fj,Pouthier:2004fk,Falvo:2006ly} It is these couplings that are responsible for the appearance of two bound states in the polaron energy spectrum as seen in Fig.~\ref{fig:energy_ac}.
\subsubsection{Linear absorption}
Using Eqs.~(\ref{eq:Jt}) and~(\ref{eq:Ck}) one can easily show  that in the limit $N\rightarrow \infty$, the linear optical response is written
\begin{equation}
J(t) = N\mu^2 \sum_{n} e^{-\ii \tilde{\omega}_0 t} e^{-g_n(t)} (-\ii)^n J_n\left(2\tilde{J}(\beta) t\right),
\label{eq:Jtac}
\end{equation}
where $J_n(x)$ are the Bessel functions of the first kind. %
For the case of a vanishing hopping constant $J=0$, i.e. the anti-adiabatic limit, the optical response is written
\begin{equation}
J(t) = N\mu^2 \exp\left(-\ii\tilde{\omega}_0 t -g_0(t)\right),
\end{equation} 
where the linebroadening function $g_0(t)$ is written
\begin{equation}
g_0(t) = \frac{4S_{\text{ac}}}{\pi} \int_{0}^{1} \frac{\ud x}{x} \sqrt{1-x^2} \left\{\coth \left(\beta\Omega_c x/2 \right) (1-\cos\Omega_c t x) +\ii \sin \Omega_c t x \right\}.
\end{equation}
Next, two situations are considered: the case of the strong coupling limit where $S_{\text{ac}} \gg 1 $ and the case of weak coupling limit $S_{\text{ac}} \ll 1$.\par
In the strong coupling limit $S_{\text{ac}}\gg 1$, the dynamics is controlled by the behavior of the correlation function at short times. Therefore, a second-order Taylor expansion of the linebroadening function can be performed in the time variable $\tau = \Omega_c t$, giving
\begin{equation}
g_0(\tau) = \ii S_{\text{ac}} \tau + S(\beta)\tau^2 / 4  + \mathcal{O}(\tau^3).
\end{equation}
The absorption spectrum is then Gaussian
\begin{equation}
\alpha(\omega) = N\mu^2 \sqrt{\frac{4\pi}{S(\beta)\Omega_c^2}} \exp\left(-\frac{(\omega-\omega_0)^2}{S(\beta)\Omega_c^2}\right).
\label{eq:gaussian}
\end{equation}
This expression is valid for all temperature range.\par
\begin{figure}
\includegraphics{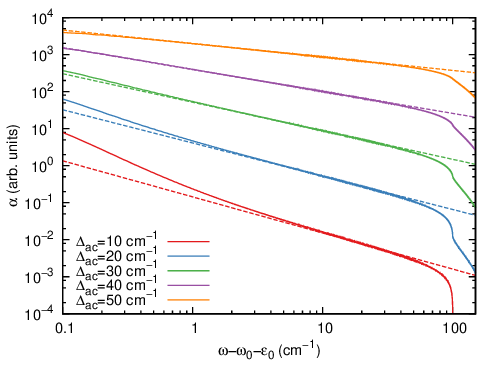}
\caption{Linear absorption spectrum as a function of the phonon coupling $\Delta_{\text{ac}}$ constant for $\Omega_{\text{ac}}=~100~\cm$, $J=0~\cm$, $T= 0$ K. To show the specific effect of the phonon broadening the relaxation time was increased to $T_1=50$ ps. The dashed line corresponds to the absorption spectrum computed from Eq.~(\ref{eq:linac}).}
\label{fig:pow_ac}
\end{figure}

In the weak coupling limit $S_{\text{ac}}\ll 1$, the dynamics is controlled by the behavior of the correlation function at long times which strongly depend on the temperature range.\cite{Duke:1965cr} For the  high temperature case $\beta\Omega_c \ll 1$ the linebroadening function is written
\begin{equation}
g_0(\tau) = \frac{8S_{\text{ac}}}{\pi\beta\Omega_c} \int_{0}^{1} \frac{\ud x}{x^2} \sqrt{1-x^2} (1-\cos \tau x),
\end{equation}
which gives an analytical but cumbersome expression as a function of the Bessel functions and the Struve functions.\cite{Abramowitz:1972qf} At long timescales one can use the continuum approximation which only consider the effect of the low-frequency phonons $x = \Omega_q/\Omega_c \sim q / 2\rightarrow0$. With this approximation the linebroadening function is written
\begin{equation}
g_0(\tau) = \frac{4S_{\text{ac}}}{\beta\Omega_c} | \tau | , 
\end{equation}
The linebroadening function therefore increases linearly with time when $\tau\rightarrow\infty$. The absorption spectrum has a Lorentzian shape with a width proportional to the temperature and shifted by the bath reorganizational energy
\begin{equation}
\alpha(\omega) = N\mu^2 \frac{8S_{\text{ac}}/\beta}{\left( \omega -\omega_0 - \epsilon_0 \right)^2 + \left(4S_{\text{ac}}/\beta\right)^2 }.
\end{equation} 
In the low temperature regime $\beta\Omega_c \gg 1$, the linebroadening function is written as
\begin{equation}
g_0(\tau) = \frac{4S_{\text{ac}}}{\pi} \int_{0}^{1} \frac{\ud x}{x} \sqrt{1-x^2} \left( 1 - e^{-\ii \tau x } \right).
\end{equation}
An asymptotic expansion of this expression gives 
\begin{equation}
g_0(\tau) = \frac{4S_{\text{ac}}}{\pi}\left( \gamma - 1 + \log 2 + \log \tau \right) + 2\ii S_{\text{ac}} \sgn \tau,
\end{equation}
where $\sgn \tau$ is the sign function and where $\gamma \approx 0.577216$ is  Euler's constant. The linebroadening function increases logarithmically with time. Introducing the coupling constant $\delta=4S_{\text{ac}}/\pi$, for a weak coupling $\delta<1$ the Fourier transform of the function $J(t)$ can be calculated and the absorption spectrum is then given by a power law for $\omega>\tilde{\omega}_0=\omega_0-\epsilon_0$ 
\begin{equation}
\alpha(\omega) = \frac{2 \sin(\delta \pi ) \Gamma(1-\delta) e^{-\delta(\gamma - 1)}}{ (2\Omega_c)^{\delta} (\omega- \tilde{\omega}_0)^{1-\delta}} \Theta(\omega - \tilde{\omega}_0)
\label{eq:linac}
\end{equation}
where $\Gamma(x)$ is the gamma function. Fig.~\ref{fig:pow_ac} shows a direct comparison of the absorption spectrum at $T=0$ K computed directly from the linear response function with the approximation of Eq.~(\ref{eq:linac}). The spectrum was centered around its maximum located at $\tilde{\omega}_0=\omega_0 - \epsilon_0$. To highlight the effect of the phonon broadening, the lifetime $T_1$ was increased to 50 ps. For a coupling constant $\Delta_{\text{ac}}=50~\cm$, the long time approximation perfectly match the full calculation for frequencies lower then 50~\cm. As decreasing the coupling constant $\Delta_{\text{ac}}$, discrepancies occur around the maximum of the absorption band corresponding to long time behavior from the response function. This originate from the ad-hoc relaxation $T_1$ introduced in the numerical calculation  which is not included in Eq.~(\ref{eq:linac}). Note that in the low temperature regime, the effect of the remaining coupling neglected in this work might control the spectral bandwidth.\par
\begin{figure}
\includegraphics{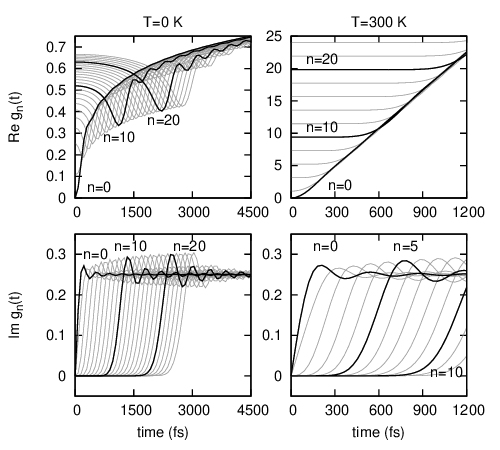}
\caption{Linebroadening function $g_n(t)$ computed from the acoustical phonon model for $\Omega_{\text{ac}}=~100~\cm$ and $\Delta_{\text{ac}}=25~\cm$ as a function of the time $t$ and distance $n$ for $T=0$ K and $T=300$ K.}
\label{fig:gfct_ac}
\end{figure}

For $J\neq 0$, effects from the polaron dispersion should arise through the sum over the index $n$  in Eq.~(\ref{eq:Jtac}). However as $n$ increases the linebroadening function increases quickly and its contribution decreases in the expression of the response function $J(t)$. For example, in the weak-coupling limit, at high-temperature and using the continuum approximation the linebroadening function is written %
\begin{equation}
g_n(\tau) = \frac{2S_{\text{ac}}}{\beta\Omega_c} \left(  | \tau  - 2 n | + | \tau + 2n | \right),
\end{equation}
which increases linearly with $n$. In this limit, the contribution of the polaron bandwidth is then negligible. In the low-temperature limit, $g_n(0)$ increases with $\log n$ and only small effects of $J$ on the absorption spectrum should be expected. Fig.~\ref{fig:gfct_ac} shows the linebroadening function $g_n(t)$ as a function of time $t$ and distance $n$ in the low temperature regime ($T=0$ K) and in the high temperature regime ($T=300$ K). In the low temperature regime, the real part of the linebroadening function shows clearly a logarithmic behavior for all $n$ at long timescales. The imaginary part has a step function behavior. In the high-temperature limit, the real part of the linebroadening function is constant for $\Omega_c t < 2 m$ and then increases linearly with time.
 \par
\subsubsection{Nonlinear spectra}
To simplify, the case $J=0$ is considered. Since acoustical phonons  are considered, even if the sites are not directly coupled in this limit through the hopping constant, coupling with the phonon bath can introduce long-range phonon-mediated correlations. For $J=0$, the sum of  GSB and ESE response functions is given by
\begin{equation}
R_1+ R_2  = N\mu^4 e^{\ii \tilde{\omega}_0(t_1-t_3)}\sum_m \left( e^{-g^{(1)}_{0, m, 0}} + e^{-g^{(2)}_{m, -m, m}} \right).
\label{eq:J0a}
\end{equation}
If there is no coupling to the bath, this contribution to the nonlinear signal scales as $N^2$. For a weak coupling with the bath, the scaling of this contribution will strongly depend on the temperature. In the low temperature regime, the linebroadening function scales as $g_m \approx \delta \log m$. In this case, the sum of the ESE and GSB contributions  scales as $N^{2-\delta}$. At high-temperature, the linebroadening function scales as $g_m \sim m$ and the sum of the ESE and GSB contributions scales as $N$. The ESA response function is written as
\begin{align}
R_3 & =  2 N\mu^4 e^{\ii \tilde{\omega}_0(t_1-t_3)}  \left(e^{-2 \ii (\epsilon_0 + A)t_3} -1 \right) e^{-g^{(3)}_{0,0,0}}  \nonumber \\ 
&  +  2N\mu^4  e^{\ii \tilde{\omega}_0(t_1-t_3)}  \left(e^{-2\ii \epsilon_1 t_3} -1 \right) \left( e^{-g^{(3)}_{1,-1,1}} +  e^{-g^{(3)}_{0,1,0}}\right) \nonumber \\
& + N\mu^4e^{\ii \tilde{\omega}_0(t_1-t_3)} \sum_m \left( e^{-g^{(3)}_{0, m, 0}} + e^{-g^{(3)}_{m, -m, m}} \right).
\label{eq:J0b}
 \end{align}
 The first two terms scale as $N$ while the last term is almost identical to the sum of the ESE and GSB contributions. In fact this term completely compensates Eq.~(\ref{eq:J0a}) if $g^{(1)}_{0,m,0} = g^{(3)}_{0,m,0}$ and  $g^{(2)}_{m, -m, m} = g^{(3)}_{m, -m, m}$. In this case, the spectra is given by one positive peak located on the diagonal and two negative peaks shifted due to the two types of anharmonicity. However if $g^{(1)}_{0,m,0} \neq g^{(3)}_{0,m,0}$ or  $g^{(2)}_{m, -m, m} \neq g^{(3)}_{m, -m, m}$   this is not true anymore. In fact for the ESE contribution the difference can be expressed as
 \begin{align}
e^{-g^{(2)}_{m,-m,m}} -  e^{-g^{(3)}_{m,-m,m}} &\propto e^{-g_m^*(t_3)} - e^{-g_m(t_3)} \nonumber \\
   & \propto \ii \sin(g''_m(t_3)),
\label{eq:diffg}
 \end{align}
 where $g''_m(t)$ is the imaginary part of the linebroadening function. The imaginary part of the linebroadening function in the long time approximation is given by
 \begin{equation}
 g''_m(t_3) \approx 2 S_{\text{ac}} \sgn(\Omega_c t_3 - 2 m),
 \end{equation} 
and do not vanish if $\Omega_c t_3$ is larger than $2m$. Note that the difference appears with a $\pi/2$ dephasing which originates from the complex factor $\ii=\sqrt{-1}$ in Eq.~(\ref{eq:diffg}). Consequently, this contribution to the spectrum is therefore dispersive and results in a pair of positive and negative peaks located on the diagonal and along the $\omega_3$ axis  as noticed in Fig.~\ref{fig:2d_ac_2}. This effect however will decrease quickly as temperature increases  and the lineshape function decay faster. It can only be seen at very low temperature as observed in Fig.~\ref{fig:2d_ac_2}.\par
\begin{figure}
\includegraphics{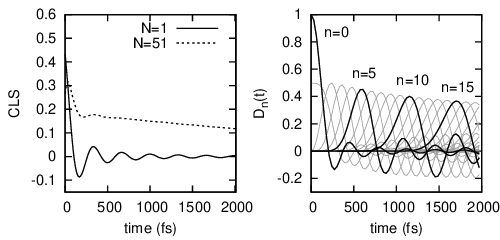}
\caption{Left panel: Center line slope of the 2D-IR spectra for the acoustical model for $\Omega_{\text{ac}}=~100~\cm$,  $\Delta_{\text{ac}}=25~\cm$ and $T=300$ K, $J=-10$~\cm\  as a function of the delay time $t_2$, for a 1D chain (dashed line) and for a single isolated site $N=1$ (full line). Right panel: bath correlation functions $D_n(t)$ as a function of time.}
\label{fig:correl_ac}
\end{figure}

On Fig.~\ref{fig:2d_ac_1}, strong differences have been observed between the case of an isolated site $N=1$ and the case of a 1D chain coupled to an acoustical phonon bath. These differences originate mostly from the bath mediated correlations. To quantify the time evolution of the 2D spectrum, the CLS has been computed. The CLS is commonly used as a metric to quantify the fluctuation timescales and extract the FFCF from 2D spectra.\cite{Kwak:2007fk,Kwak:2008uq,Roy:2011xe,Falvo:2016fk} The CLS for the isolated site and the 1D chain are represented on the left panel of Fig.~\ref{fig:correl_ac}. For $N=1$ the CLS decays quickly over the first 200 fs with an oscillation corresponding to the frequency $\Omega_c$. This behavior is characteristic of an underdamped Brownian oscillator. For the 1D chain, the CLS still exhibits a fast decay over the first 200 fs but then it decays on a much slower timescale. A convenient way to interpret these results is to introduce the FFCF for a delocalized system
\begin{equation}
D_n(t) = \langle \hat{v}_n(t) \hat{v}_0(0) \rangle, 
\end{equation}
where $\hat{v}_n = \frac{1}{\sqrt{N}}  \sum_{q} ( \Delta_{q} e^{-\ii q n} a_{q}^\dagger + \Delta_{q}^{*} e^{\ii q n} a_q )$ is the frequency fluctuation operator of site $n$. This  function measures the correlation of the fluctuations between two sites separated by the distance $n$ and after the time $t$. It  can be written as
\begin{equation}
D_n(t) = \frac{1}{N} \sum_q |\Delta_q|^2 \left( \coth \left(\frac{\beta \Omega_q}{2}\right) \cos(\Omega_q t - q n)   -\ii \sin( \Omega_q t - q n )\right).
\end{equation}
For the acoustical phonon model, in the high temperature limit and for $N\rightarrow\infty$, the bath correlation function is written as a function of the variable $\tau=\Omega_c t$ as
\begin{equation}
D_n(\tau) = \frac{16 \Delta_{\text{ac}}^2}{\pi \beta\Omega_c} \int_0^1 \ud x \sqrt{1-x^2} \cos(\tau x) T_{2n}\left(\sqrt{1-x^2}\right),
\end{equation}
where $T_m(x)$ are the Chebyshev polynomials of the first kind.\cite{Abramowitz:1972qf} For example for $n=0,1,2$, the first three functions $D_n(\tau)$ are written
\begin{align}
&D_0(\tau) = \frac{8 \Delta_{\text{ac}}^2}{\beta\Omega_c} \frac{J_1(\tau)}{\tau},\\
&D_1(\tau) = \frac{8 \Delta_{\text{ac}}^2}{\beta\Omega_c} \left(  \frac{6J_2(\tau)}{\tau^2} - \frac{J_1(\tau)}{\tau}   \right), \\
&D_2(\tau) = \frac{8 \Delta_{\text{ac}}^2}{\beta\Omega_c} \left(  \frac{J_1(\tau) ( \tau^2 - 120 ) }{\tau^3}  - \frac{24J_2(\tau)(\tau^2-20)}{\tau^4}   \right).
\end{align}
For larger distances $n$ the correlation function can be in principle calculated analytically but the corresponding expressions are cumbersome and involve polynomials in $\tau$ of order $n$. The FFCFs are represented on the right panel of Fig.~\ref{fig:correl_ac}. For $n=0$, the correlation function decays in a similar manner as the CLS for $N=1$ with the same underdamped oscillation. The CLS appears just shifted compared to the FFCF. This is not a surprise because it has been shown that by including fast fluctuation processes the CLS measures directly the scaled and shifted FFCF.\cite{Falvo:2016fk} As $n$ increases the maximum of the FFCF is then located at $\Omega_c t = n$ and then decreases quickly with underdamped oscillations. Looking at the maximum value of the FFCF as a function of $n$ one can see that this maximum decreases slowly in a similar fashion as the CLS for the 1D chain. This shows that the bath-mediated long-range correlations measured by the FFCFs is directly responsible for the slow decay of the CLS in the 2D-IR spectrum.\par
\subsubsection{Experimental implications}
In Ref.~\citenum{Edler:2004qy}, Edler and coworkers measured the pump-probe spectrum of a model $\alpha$-helix in the NH spectral range. They observed two bound states that were interpreted as the signature of the two anharmonic couplings from Eq.~(\ref{eq:anharcouplings}): one anharmonic coupling that create a pair of excitons located on the same site and one anharmonic coupling that creates a pair of excitons located on two nearest-neighbor sites. A similar observation was then made by Bodis and coworkers on a model $\beta$-sheet peptide.\cite{Bodis:2009uq}  The model used to interpret the experiment of Ref.~\citenum{Edler:2004qy}, relied on a slightly different Hamiltonian than the one presented in this paper as it included also the fluctuations of the anharmonicity due to the phonons. To explain the experimental observations it was assumed a very strong coupling between the NH vibrations and the accoustical phonon bath, which would correspond with the present model to a coupling constant of $S_{\text{ac}}=1.2$. However, the intrepretation used in Ref.~\citenum{Edler:2004qy} did not take into account the spectral broadening induced by the bath on the linear and the non-linear spectra. In the high temperature limit, the dressing factor is given by $S(\beta) = 4 S_{\text{ac}} k_{\text{B}}T / \Omega_{\text{ac}}$. Using Eq.~(\ref{eq:gaussian}) we can compute the FWHM of the linear absorption spectrum in the strong coupling limit. It is given by
\begin{equation}
\Delta \omega = 4 \sqrt{S_{\text{ac}}\Omega_{\text{ac}} k_{\text{B}}T \ln 2}
\end{equation}
Therefore the FWHM is proportional to the square root of the temperature. For a temperature of 300 K and a coupling constant $S_{\text{ac}}=1.2$ the FWHM is $525~\cm$. This value, in complete disagreement with the measurements, demonstrates that the model is unable to explain all the experimental observations.  In addition, as was observed in Ref.~\citenum{Edler:2004qy}, the spectral bandwidth in the NH range do not change significantly with temperature. This shows that the Davydov Hamiltonian cannot explain the emergence of two bound states in the NH pump-probe spectra of $\alpha$-helices and $\beta$-sheets. Therefore additional theoretical work is needed to explain these bound states.
%
\section{Conclusions}
\label{sec:conclusions}
In this article, a new  methodology to calculate the non-linear response of vibrational systems based on the small polaron approach is presented. This approach relies on a unitary transformation which dresses the vibrational excitation by a phonon cloud and is valid in the anti-adiabatic limit where the hopping constant is small with respect to the phonon frequency. This method allows to calculate the optical response of large systems and can describe explicitly bath-mediated correlation. This method was used to calculate the linear and non-linear spectroscopy of 1D model chains considering both optical and acoustical phonon bath.\par
For the case of an optical phonon bath, a simple expression was derived for the linear absorption. It shows that the absorption spectrum is characterized by a main zero-phonon line and a series of $n$th-phonon lines. The $n$th-phonon lines lineshape is given by the polaron density of states. Here, the case of a 1D lattice has been considered  and the density of states is characterized by a double peak. This result can be transposed to the more complicate case of an $n$-dimensional (nD) lattice. The density of states will strongly depend on the dimensionality of the problem and therefore impact the shape of the absorption spectrum. This confirms the result of Hamm and Edler\cite{Hamm:2006uq} which states that 3D effects might play an important role to explain linear absorption spectrum of ACN. The approach presented in this article represents a first step towards a full understanding of the linear and non-linear spectra of crystaline acetanilide. It will be extended to include explicitly the 3D structure of ACN allowing therefore a direct comparison with experiment.\par
For the case of the acoustical phonon bath, this article shows that in the C$=$O spectral range, two bound states can be clearly visible in the 2D-IR spectrum at low temperature. However only one is visible at ambiant temperature. Up to now, no experimental measurements of 2D-IR spectroscopy of $\alpha$-helix polypeptides have been made at very low temperature in the C$=$O spectral range. New measurements in this temperature range could therefore bring new information on the vibrational dynamics in this system. This article shows that for the case of acoustical phonons at ambiant temperature, the spectral diffusion measured from the 2D-IR spectrum appears much slower for the case of a lattice than for a single site. This can be explained by bath mediated correlations between distant sites. Experimental observation of such process could be performed on a model $\alpha$-helix for example by measuring the 2D-IR spectrum of the $\alpha$-helix and compare it to an isotope labeled $\alpha$-helix for which the spectrum of a single isolated vibration coupled to the full phonon bath could be obtained. Finally, the present model puts in question the validity of the model used to explain the pump-probe spectra measured in the N$-$H spectral range for a model $\alpha$-helix and a model $\beta$-sheep peptides. The model predicts a spectral bandwidth that increases with the square root of the temperature, a behavior in disagreement with the experimental measurements.\par
The main limitation of this work resides in the fact that the residual coupling between the polaron and the bath has been neglected. This coupling can modify the absorption spectrum and the 2D-IR spectrum. Further theoretical developments are necessary to fully account for this coupling in the linear and non-linear response. In particular, inclusion of the remaining coupling by perturbation theory seems a very promising approach.\cite{Pouthier:2004fk,Pouthier:2013fk,Yalouz:2016aa}
\begin{acknowledgments}
The author gratefully acknowledges financial support by the Agence Nationale de la Recherche (ANR) grants ANR-11-BS04-0027 and ANR-16-CE29-0025 as well as the use of the computing center M\'esoLUM of the LUMAT research federation (FR LUMAT 2764).
\end{acknowledgments}
\appendix
\section{Optical and Acoustical phonon models}
\label{app:phonon}
In the optical phonon model, the bath is characterized by a set of vibrational coordinates $u_n$ and corresponding momentum $p_n$ and of harmonic frequency $\Omega_{\text{opt}}$. The bath Hamiltonian is then written as
\begin{equation}
\hat{H}_b = \sum_{n} \frac{p_n^2}{2} + \frac{ \Omega_{\text{opt}}^2}{2} u_n^2.
\end{equation} 
I will consider that the exciton frequency of site $n$ is linearly coupled to the $n$th bath mode. Therefore the system-bath coupling Hamiltonian is written
\begin{equation}
\hat{H}_{vb} = \sum_{n} \chi u_n b_n^\dagger b_n,
\end{equation}
where $\chi$ is a coupling constant. Using a  plane wave basis and  the phonon creation and annihiliation operators the expression for the bath coordinates $u_n$ is given by
\begin{align}
u_n = \frac{1}{\sqrt{2N\Omega_{\text{opt}}}} \sum_q \left( e^{\ii q n} a_q + e^{-\ii q n} a_q^\dagger \right).
\end{align}
Similarly the corresponding momentum $p_n$ is written
\begin{equation}
p_n = \ii\sqrt{\frac{\Omega_{\text{opt}}}{2N}} \sum_q \left( e^{-\ii q n} a^\dagger_q - e^{\ii q n} a_q \right).
\end{equation}
Using these expressions, one can immediately obtain the expression for the bath and coupling Hamiltonians Eqs.~(\ref{eq:Hb}) and~(\ref{eq:Hvb}) with $\Omega_{q} = \Omega_{\text{opt}}$ and $\Delta_q = \chi/\sqrt{2\Omega_{\text{opt}}}$.\par
For the model of acoustical phonons, the bath Hamiltonian is written as a function of the bath mode $u_n$ as
\begin{equation}
\hat{H}_b = \sum_{n} \frac{p_n^2}{2} + \frac{W}{2}\left( u_{n+1} - u_{n} \right)^2,
\end{equation}
where $W$ is a coupling constant. Following Davydov,\cite{Davydov:1985aa} the system-bath coupling Hamiltonian is written as
\begin{equation}
\hat{H}_{vb} = \sum_n \chi \left( u_{n+1} - u_{n-1} \right) b_n^\dagger b_n.
\end{equation}
Using the phonon creation and annihilation operators, the expression for the bath coordinates and momentum are given by
\begin{align}
u_n =  \sum_q \frac{1}{\sqrt{2N\Omega_q}} \left( e^{\ii q n} a_q + e^{-\ii q n} a_q^\dagger \right), \\
p_n =  \sum_q \ii\sqrt{\frac{\Omega_q}{2N}} \left( e^{-\ii q n} a^\dagger_q - e^{\ii q n} a_q \right),
\end{align}
where $\Omega_q = \Omega_{\text{ac}} | \sin q / 2|$ and where the cutoff frequency is given by $\Omega_{\text{ac}} = \sqrt{4W}$. One can then obtain immediately the expression for the coupling Hamiltonian Eq.~(\ref{eq:Hvb}) with
\begin{equation}
\Delta_{q} = -2 \ii \chi W^{-1/4} \sqrt{|\sin q/2|} \cos q/2.
\end{equation} 
\section{$\textbf{k}_{\text{II}}$ response functions}
\label{app:signals}
The signal in the direction $\textbf{k}_{\text{II}}$ can be written as the sum of three contributions
\begin{equation}
R_{\textbf{k}_{\text{II}}}(t_1,t_2,t_3) = R_4 + R_5 - R_6,
\end{equation}
where $R_4$ and $R_5$ are respectively the GSB and ESE while $R_6$ is the ESA. These response functions are written
\begin{align}
R_4(t_1,t_2,t_3) &= \frac{\mu^4}{N} \sum_{k_1 k_2} e^{-\ii \omega_{k_1} t_1 -\ii \omega_{k_2} t_3 } C^{(4)}_{k_1 k_2}(t_1,t_2,t_3) \\
R_5(t_1,t_2,t_3) &= \frac{\mu^4}{N} \sum_{k_1 k_2} e^{ -\ii \omega_{k_1} (t_1+t_2+t_3)+ \ii \omega_{k_2} t_2 } C^{(5)}_{k_1 k_2}(t_1,t_2,t_3) \\
R_6(t_1,t_2,t_3) &= \frac{\mu^4}{N^2} \sum_{k_1 k_2 k_3 \sigma} e^{ -\ii \omega_{k_1} (t_1+t_2) +  \ii \omega_{k_2} (t_2+t_3) -\ii \omega_{k_3 \sigma} t_3 } C^{(6)}_{k_1 k_2 k_3}(t_1,t_2,t_3)A_{k_1 k_3 \sigma} A_{k_2 k_3 \sigma}
\end{align}
with the functions $C^{(i)}(t_1,t_2,t_3)$ defined by
\begin{align}
C^{(4)}_{k_1 k_2} (t_1,t_2,t_3) &=  \sum_{m_1 m_2 m_3} e^{\ii k_1 m_1 + \ii k_2 m_3} e^{-g^{(4)}_{m_1 m_2 m_3}(t_1,t_2,t_3) } \\
C^{(5)}_{k_1 k_2} (t_1,t_2,t_3) &=  \sum_{m_1 m_2 m_3} e^{\ii k_1 (m_1+m_2+m_3) - \ii k_2 m_2} e^{-g^{(5)}_{m_1 m_2 m_3}(t_1,t_2,t_3) } \\
C^{(6)}_{k_1 k_2} (t_1,t_2,t_3) &= \sum_{m_1 m_2 m_3} e^{ \ii k_1 (m_1+m_2) -\ii k_2 (m_2+m_3) + \ii k_3 m_3} e^{-g^{(6)}_{m_1 m_2 m_3}(t_1,t_2,t_3) }
\end{align}
and where the linebroadening functions are given by
\begin{align}
&g^{(4)}_{m_1m_2m_3}(t_1,t_2,t_3) = g_{m_1}(t_1) +g_{m_2}(t_2) +  g_{m_3}(t_3) \nonumber \\ 
     &- g_{m_1+m_2}(t_1+t_2) -  g_{m_2+m_3}(t_2+t_3)  +g_{m_1+m_2+m_3}(t_1+t_2+t_3) \\
&g^{(5)}_{m_1m_2m_3}(t_1,t_2,t_3) = g_{m_1}(t_1) +g^*_{m_2}(t_2) +  g^*_{m_3}(t_3) \nonumber \\ 
     &- g_{m_1+m_2}(t_1+t_2) -  g^*_{m_2+m_3}(t_2+t_3)  +g_{m_1+m_2+m_3}(t_1+t_2+t_3) \\
&g^{(6)}_{m_1m_2m_3}(t_1,t_2,t_3) = g_{m_1}(t_1) +g^*_{m_2}(t_2) +  g_{m_3}(t_3) \nonumber \\ 
     &- g_{m_1+m_2}(t_1+t_2) -  g^*_{m_2+m_3}(t_2+t_3)  +g_{m_1+m_2+m_3}(t_1+t_2+t_3) 
\end{align}
\bibliography{biblio}
\end{document}